\documentclass[twocolumn,prl,showpacs]{revtex4}

\usepackage{graphicx}

\begin{document}
\title{Quantum measurement theory for particle oscillations}

\author{Charis Anastopoulos}
\email{anastop@physics.upatras.gr} \affiliation{Department of
Physics, University of Patras, 26500 Patras, Greece}
\author{ Ntina Savvidou} \email{ntina@imperial.ac.uk}
 \affiliation{  Theoretical Physics Group, Imperial College, SW7 2BZ,
London, UK}


\begin{abstract}
A fundamental principle  of quantum theory, clearly manifested in
the two-slit experiment, is that for any alternatives that cannot be
distinguished by measurement physical predictions are obtained by
summation of  their amplitudes. In particle oscillation experiments,
a particle's time of detection is not directly measured,
consequently, the detection probability should involve the summation
over amplitudes corresponding to different detection times. However,
in contrast to the principle above, standard treatments involve
summation {\em over probabilities} rather than amplitudes; this
implicitly assumes the existence of a decohering mechanism. In this
work, we construct the detection probabilities for particle
oscillations by summation {\em over amplitudes}, corresponding to
different detection times. The resulting wavelength of particle
oscillations differs from the standard expression by a factor of
two. Moreover, we predict a dependence of the oscillation wavelength
on the threshold of the decay process used for detection.

\end{abstract}

\maketitle

The two-slit and related experiments demonstrate a striking feature
of quantum mechanics: for any two alternatives that cannot be
distinguished by a measurement scheme, their joint probability is
obtained by summing their contributions at the level of {\em
probability amplitudes}; therefore  interference terms appear. If,
however, a macroscopic distinction of these alternatives is
possible, the interference terms are suppressed and the joint
probability is the sum of the individual probabilities.

Let us examine the importance of the above elementary point for the
correct quantum treatment of  particle oscillations \cite{osc}. The
detection time of the oscillating particles is not directly
measured. Instead, we measure the number of detection events, at
distance $L$ from the source. The question then arises: how,
starting from the probability amplitude ${\cal A}_{\alpha}(t,
x)$---for the detection of a flavor $\alpha$, at a point $x$, at
sharply defined moment of time $t$---, we can obtain the probability
$p_{\alpha}(L)$ that a particle of flavor $\alpha$ will be detected
at distance $L$ from the source.

A common approach is to substitute in ${\cal A}_{\alpha}(t, x)$ the
time $t$ with the classical time of arrival at $L$, namely $L/c$ for
neutrinos, and to write the corresponding probability as
$p_{\alpha}(L) \sim |{\cal A}_{\alpha}(L, L)|^2$. The usual setting
for this approach is the textbook plane-wave treatment of neutrino
oscillations, where
\begin{eqnarray}
{\cal A}_\alpha(t, x) = \sum_{i}U^*_{\beta i}U_{\alpha i}\ e^{ i
p_{i}x - iE_i t }. \label{ampl00}
\end{eqnarray}
Here, $i$ labels the mass eigenstates of neutrinos, $U_{i \alpha}$
is the mixing matrix, $\beta$ is the initial flavor, $p_i$ are the
momenta in the different mass-eigenspaces and the corresponding
energies $E_{i} = \sqrt{m_i^2 +p_i^2}$. The substitution  $t = L$ in
Eq. (\ref{ampl00}) and the evaluation of the detection probability
yields the standard expression for the oscillation wavenumber $
k_{ij} = (m_i^2 - m_j^2)/(2p)$, for $p_i = p +O(m_i)$.

This method involves an unsatisfactory mixing of classical and
quantum concepts and this problem persists if we consider realistic
wave packets rather than plane waves. More importantly, the method
is ambiguous in the following way: different mass eigenstates
correspond to different velocities $v_{i} = p_i/E_{i}$; if we
evaluate each mass-component at the corresponding arrival times
$t_{i} = L/v_{i}$, then we obtain an oscillation wave-number $k_{ij}
= (m_i^2 - m_j^2)/p$ for $p_i = p +O(m_i)$, which is twice the
standard result \footnote{For derivations of non-standard
expressions for the oscillation wavelength, see Ref. \cite{Beu03}
and references therein.}.

Furthermore, the standard method assumes that the detection time
should be computed in terms of an `average' velocity  for all
components of the wave packet---see, for example, Refs. \cite{Giu03,
Lip}. This assumption involves the substitution of a coherent
superposition by an altogether different state that cannot be done
in an invariant way in quantum mechanics. Any choice of an `average
velocity' for a wave packet is arbitrary and unjustified at a
fundamental level.


An improved derivation of the standard expression for the
oscillation wavelength proceeds by calculating the time average of
probabilities over detection time, i.e.,  defining $p_{\alpha}(L)
\sim \int_0^T dt |{\cal A}_{\alpha}(t, L)|^2$, where $T$ is the
integration time for the experiment. However, a summation of
probabilities over detection time is not justified by the rules of
quantum theory. As the two-slit experiment indicates, alternatives
that cannot be distinguished by measurement---in this case,
detections of a flavor $\alpha$ at different moments of time---are
to be summed over at the level of amplitudes.

The summation over amplitudes, defined at different moments of
 time, is rather intricate because there is no time
operator in quantum theory. Hence the methods that apply to other
observables are not directly applicable here. The method we present
in this letter has the following advantages: (i) it fully implements
the basic principle, that alternatives that are not distinguished by
measurement are to be summed at the level of amplitudes; (ii) it
incorporates a genuinely quantum treatment of detection time,
according to the quantum theory of measurement \cite{QMT}; (iii) it
employs a general formula, Eq. (\ref{prob}), that is derived using
only the rules of quantum theory.  Eq. (\ref{prob}) is valid for
{\em any} setup in which the detection time is not measurable, not
only to particle oscillations. The end result is an expression for
the detection probability as a function of distance; it leads to the
 {\em non-standard} formula for the oscillation
wave-number $k_{ij} = (m_i^2 - m_j^2)/p$ that was mentioned
previously.

In what follows, we describe in more detail our approach and the
derivation of the results above. We employ a method developed in
\cite{AS},  for the construction of probabilities from amplitudes
that are defined at different moments of time. It has been applied
to various problems, such as probabilities for time-of-arrival,
tunneling-time and for non-exponential decays. This method contains
ideas from the decoherent histories approach to quantum mechanics
\cite{scg} and it has many similarities to the Srinivas-Davies
photo-detection theory \cite{SD81}.

The first step in our study is the derivation for a general formula
for the probability of detection outcomes, when the time of
detection is not observable. (The presentation here is simplified:
for elaboration on the finer points, see Ref. \cite{AS}.) Let ${\cal
H}$ be the Hilbert space of a quantum system, and $\hat{H}$ the
Hamiltonian operator. In order to describe an event---such as a
particle detection---we introduce a projection operator $\hat{P}$
that corresponds to states
 accessible only if the event has taken place. For example,
if we detect a particle by monitoring a specific decay process, then
$\hat{P}$ is a projector onto the states of all product  particles.

The Hilbert space ${\cal H}$ splits into two subspaces corresponding
to the projectors $\hat{P}$ and $\hat{Q} = 1 - \hat{P}$. Let
$\hat{P}_\lambda$ be the projection operators corresponding to
different values $\lambda$ of an observable that can be measured
only if a detection event has occurred. For example,
$\hat{P}_\lambda$ may correspond to a coarse-grained position
variable for a product particle. The set of projectors
$\hat{P}_\lambda$ is exclusive ($\hat{P}_{\lambda}
\hat{P}_{\lambda'} = 0, $ if $\lambda \neq \lambda'$) and
exhaustive, provided a detection has occurred; i.e., $\sum_\lambda
\hat{P}_\lambda = \hat{P}$.

 Starting from the initial state of  a quantum system $|\psi_0
\rangle$, we construct the amplitude $|\psi; t, \lambda \rangle$, at
 a final time $T$,  that arises if a detection event took place within the
interval $[t, t+ \delta t] \subset [0, T]$ and the outcome of the
measurement was $\lambda$. By assumption, no detection took place
before time $t$. Hence, we evolve $|\psi_0 \rangle$ with $\hat{S}_t
= \lim_{N \rightarrow \infty} (\hat{Q}e^{-i\hat{H} t/N} \hat{Q})^N$,
i.e., with the restricted propagator in the eigenspace of the
projector $\hat{Q}$ to no-detection states. In the interval $[t,
t+\delta t]$ the state evolves according to the full unitary
evolution $e^{-i \hat{H} \delta t}$. Then, we act by a projection
$\hat{P}_{\lambda}$ that selects the measurement  outcome $\lambda$.
Next, the state evolves unitarily until time $T$ because there is no
other event that needs to be taken into account. At the limit of
small $\delta t$, the successive operations above yield $| \psi; t,
\lambda \rangle = - i \, \delta t \, \,e^{-i\hat{H}(T - t)}
\hat{P}_{\lambda} \hat{H} \hat{S}_t |\psi_0 \rangle$.

We must emphasize here an important physical distinction on the role
of time pertaining to particle oscillations. The time of detection
$t$ is {\em not} identical to the evolution parameter of
Schr\"odinger' s equation. Instead, it is a {\em dynamical variable}
that determines the moment that a physical {\em event} has taken
place \cite{Sav}.
The construction of the amplitude $|\psi; t, \lambda \rangle$ above
highlights this distinction: the detection time $t$ is distinct from
the  time $T$ at which the amplitude is evaluated.

Furthermore, the amplitude $|\psi; t, \lambda\rangle$ is
proportional to $\delta t$, hence it defines  a {\em density} with
respect to time. The integration of $|\psi; t, \lambda \rangle$ over
$t$ is, therefore, well-defined in this scheme. In contrast, quantum
theory provides no natural definition of integration of single-time
probabilities over time \cite{AS}.  The amplitude $|\psi;
\lambda\rangle$ corresponding to a detection with value $\lambda$,
at {\em any} (unspecified) time $t \in [0, T]$, is
\begin{eqnarray}
| \psi; \lambda \rangle = - i \int_0^T d t \,  e^{-i\hat{H}(T - t)}
\hat{P}_{\lambda} \hat{H} \hat{S}_t |\psi_0 \rangle :=
\hat{C}_{\lambda} |\psi_0 \rangle. \label{ampl0}
\end{eqnarray}

Hence, the probability $p (\lambda)\/$ that a detection with outcome
$\lambda$ occurred at some time in $[0, T]$ is
\begin{eqnarray}
p(\lambda) = \langle \psi; \lambda | \psi; \lambda \rangle = \langle
\psi_0| \hat{C}_{\lambda}^{\dagger} \hat{C}_{\lambda}| \psi_0
\rangle \nonumber \\ = \int_0^T \,  dt \, \int_0^T dt' \; Tr
(e^{i\hat{H}( t - t')} \hat{P}_{\lambda} \hat{H} \hat{S}^{\dagger}_t
\hat{\rho}_0 \hat{S}_{t'} \hat{H} \hat{P}_{\lambda} ), \label{prob}
\end{eqnarray}
where $\hat{\rho}_0 = |\psi_0\rangle \langle \psi_0|$. The
probability measure Eq. (\ref{prob})  is positive, linear, and
normalized when the probability of no-detection $1 - \int_0^T dt
p(\lambda)$ is included.

If  the Hamiltonian is of the form $\hat{H} = \hat{H}_0 +
\hat{H}_I$,  where $[\hat{H}_0, \hat{Q}] = 0$ and $\hat{H}_I$ a
small perturbation, then, to leading order in the perturbation,
\begin{eqnarray}
p(\lambda) = \int_0^T dt \int_0^T dt' \; Tr (\hat{U}_{t' - t}
\hat{P}_{\lambda} \hat{H}_I \hat{U}_t \hat{\rho}_0
\hat{U}_{t'}^{\dagger} \hat{H}_I \hat{P}_{\lambda} ), \label{prob2}
\end{eqnarray}
where $\hat{U}_t = e^{- i \hat{H}_0 t}$.

Eqs. (\ref{prob}--\ref{prob2}) are general operator expressions
valid for any  system. Next we apply them to the case of particle
oscillations. Since the oscillating particles are detected by means
of their decay products, the precise treatment of the detection
process involves the use of quantum field theory. We will not
specify the type of oscillating particles so that our results will
be valid for both neutrino and neutral boson oscillations.

To this end, let us denote the oscillating particles as $A$ and
consider that their detection involves the process $A + B_m
\rightarrow D_n$, where $B_m$ and $D_n$ are particles labeled by
indices $m$ and $n$. The Hilbert space ${\cal H}$ of the total
system is a tensor product ${\cal H}_A \otimes {\cal H}_{r}$, where
${\cal H}_A$ is the (bosonic or fermionic) Fock space ${\cal
F}({\cal H}_{1A})$. The single-particle Hilbert space ${\cal
H}_{1A}$ is a direct sum $\oplus_{i}{\cal H}_{i}$ of mass
eigenspaces ${\cal H}_{i}$. ${\cal H}_r$ is the Hilbert space for
the degrees of freedom corresponding to the $B_m $ and $D_n$
particles. It is decomposed as ${\cal H}_0 \oplus {\cal H}_{prod}$:
${\cal H}_0$ is the subspace of states prior to the decay $A + B_m
\rightarrow D_n$ and $H_{prod}$ is the subspace corresponding to
states of the decay products \footnote{For example, if $K_L$ is
detected by means of the decay $K_L \rightarrow 3 \pi^0$, ${\cal
H}_r$ is the Fock space for $\pi^0$ particles, ${\cal H}_0$ is the
$\pi^0$ vacuum subspace and ${\cal H}_{prod}$ is the subspace of
$H_r$ with non-zero number of $\pi^0$ particles.}. Since we assume
that the measurements are carried out to the {\em product}
particles, then the projection operators $\hat{P}_{\lambda}$ in Eq.
(\ref{prob}) are of the form $1 \otimes \hat{\Pi}_{\lambda}$, where
$\hat{\Pi}_{\lambda}$ projects into a subspace of ${\cal H}_{prod}$.

We assume a Hamiltonian of the form $\hat{H} = \hat{H}_A \otimes
\hat{1} + 1\otimes \hat{H}_{r} + \hat{H}_I$, where $\hat{H}_A$ is
the Hamiltonian for the $A$ particles, $\hat{H}_{r}$ is the
Hamiltonian for the $B_m$ and $D_n$ particles, and $\hat{H}_I$ is
the interaction Hamiltonian. For simplicity, we assume that any
particles $B_m$ that are present prior to detection are stationary.
Hence,  the restriction of $\hat{H}_r$ to ${\cal H}_0$ is a
constant, which can be taken equal to zero. The restriction of
$\hat{H}_r$ on ${\cal H}_{prod}$ equals $\epsilon_{th} + \sum_n
(\sqrt{M_{D_n}^2 + \hat{{\bf p}}_n^2} - M_{D_n})$, where
$\epsilon_{th} = \sum_n M_{D_n} - \sum_m M_{B_m}$ is the threshold
energy of the $A$ particles for the process $A + B_m \rightarrow
D_n$. In the above, $M_{B_m}$ and $M_{D_n}$ are the masses of the
particles $B_m$ and $D_n$ respectively and $\hat{{\bf p}}_n$ are the
momentum operators for the $D_n$ particles.

We consider an effective interaction Hamiltonian
\begin{eqnarray}
\hat{H}_I = \sum_i \int d^3x \left[\hat{b}_i({\bf x}) U_{ \alpha i}
\hat{J}_{\alpha}^{+}({\bf x}) + \hat{b}^{\dagger}_{i}({\bf x})
U^*_{i \alpha}\hat{J}^-_{\alpha}({\bf x})\right], \label{hi}
\end{eqnarray}
 where $b_i, b^{\dagger}_i$ are annihilation and
creator operators on ${\cal H}_{A}$, $i$ labels mass eigenstates,
 $J^{\pm}_{\alpha}({\bf x})$ are current operators of flavor
$\alpha$ defined on ${\cal H}_{r}$, and $U_{i \alpha}$ is the mixing
matrix\footnote{For neutral bosons, the mixing matrix, in general,
depends on the boson's momentum; out treatment can be
straightforwardly generalized to cover this case.}. The current
operator $\hat{J}_{\alpha}^{\pm}$ involves products of annihilation
operators for the $B$ particles and creation operators for the $D$
particles. Since no $A$ particles are created during the detection
process, the initial state $|\phi_0 \rangle$ in ${\cal H}_r$ must
satisfy $\hat{J}_{\alpha}^-({\bf x}) |\phi_0 \rangle = 0$. Note that
if the detection involves a scattering process rather than a decay,
an interaction Hamiltonian quadratic to the field of the $A$
particles should be used, instead of (\ref{hi}).

Let the initial state on ${\cal H}_{A}$ be a single-particle state
$| \psi_0 \rangle = \sum_i \int d^3 x \, \hat{b}_{i}^{\dagger}({\bf
x}) \psi_{i0}({\bf x})|0 \rangle_A$, where $| 0 \rangle_A$ is the
vacuum of ${\cal H}_A$. Since $[\hat{H_0}, \hat{Q}] = 0$, Eq.
(\ref{prob2}) applies. Hence, the probability that a decay through
flavor $\alpha$ has happened at some time in $[0, T]$ and that the
value $\lambda$ for an observable of the product particles has been
found equals
\begin{eqnarray}
p_{\alpha}(\lambda) = \sum_{ij}\int_0^T dt \int_0^T dt' \int d^3x
d^3x' \psi^*_{j}({\bf x'},t') \psi_{i}({\bf x},t) \nonumber \\
\times U_{\alpha i} U^*_{j \alpha} R^{\alpha}_{\lambda}({\bf x},
{\bf x'}, t-t') , \label{prob3}
\end{eqnarray}
where $\psi_i({\bf x}, t)$ is the evolution of $\psi_{i0}({\bf x})$
under the  Hamiltonian for a single A-particle and
\begin{eqnarray}
R_{\lambda}^{\alpha}({\bf x}, {\bf x'}, t\!-\!t')\! =\!
 \langle \phi_0|\hat{J}^-_{\alpha}({\bf x'}) \hat{\Pi}_{\lambda} e\!^{-i\hat{H}_r(t'\!-t)}
  \hat{\Pi}_{\lambda} \hat{J}^+_{\alpha}({\bf x})|\phi_0\rangle.
  \label{R1}
\end{eqnarray}

We next consider the measurement of  position ${\bf X}$, of one of
the product particles, with  accuracy of order $\delta$. The
operators $\hat{\Pi}_{\lambda}$ in (\ref{R1}) can be substituted by
a Gaussian approximate projector for position
\begin{eqnarray}
\hat{\Pi}_{\bf X} = \int d^3 {\bf X'} e^{ - \frac{|{\bf X} - {\bf
X'}|^2}{ 2 \delta ^2}} |{\bf X'} \rangle \langle {\bf X'}|\otimes 1,
\end{eqnarray}
where the tensor product with unity refers to the remaining degrees
of freedom in ${\cal H}_{prod}$.

The vector $\hat{J}^+_{\alpha}({\bf x})|\phi_0 \rangle$ refers to
the state of the product particles for decays that have taken place
in a neighborhood of ${\bf x}$. Hence, the operators $\hat{\Pi}_{\bf
X}$ determine the locus of the decay event within an accuracy of
order $\delta$, if
 $\hat{\Pi}({\bf X}) \hat{J}^+_{\alpha}({\bf
x})|\phi_0 \rangle  \simeq 0 $, for $|{\bf x} - {\bf X}| >> \delta
$. Heuristically, the condition above would be satisfied for
macroscopic values of $\delta$ much larger than any length
parameters characterizing the interaction. At this level of
coarse-graining, the localization of one product particle at the
decay time essentially determines the localization of all other
product particles. In general, the localization scale $\delta$ is
macroscopic, but it has to be much smaller than the scale of
variation in the wave function of the $A$ particles, or else no
particle oscillations would be observable. Hence, to a first
approximation, we can substitute $\psi_{\alpha}({\bf x}, t) $ by
$\psi_{\alpha}({\bf X}, t) $ in Eq. (\ref{prob3}). Taking these
considerations into account, Eq. (\ref{R1}) becomes
\begin{eqnarray}
\hspace{-0.1cm} R_{{\bf X}}^{\alpha}({\bf x}, {\bf x'},\! t\!-\!t')
\! \simeq K \delta^3({\bf X}, \!{\bf x}) \delta^3({\bf X}, \!{\bf
x'}) e^{i \epsilon_{th}(t\!-\!t')} \! F(t'\!-\!t), \hspace{-0.2cm}
\label{pi2}
\end{eqnarray}
where  $K\!>\!0$  is a constant. The function $F(s)$ is obtained
from the propagator of $\hat{H}_r$ and it equals $ \prod_n \int d^3
p \; e^{ - i (\sqrt{M_{D_n}^2 + {\bf p}^2} - M_{D_n})s - \delta ^2
{\bf p}^2} $; $n$ runs over all product particles $D_n$. If $M_{D_n}
\delta
>>1$, the saddle-point approximation applies and
\begin{eqnarray}
 F(s) =\prod_n \left(\frac{ M_{D_n}}{2 \pi i (s - i M_{D_n}\delta^2/2) }\right)^{3/2}. \label{ff}
\end{eqnarray}

Eqs. (\ref{prob3}) and (\ref{pi2}) define a probability density for
the particle's position  ${\bf x}$ at scales much larger than
$\delta$
\begin{eqnarray}
p_{\alpha}({\bf x})\! =\! K \! \int_0^T\!\!\!\!\! dt
\!\int_0^T\!\!\!\!\! dt' {\cal A}_\alpha(t\!, {\bf x}) {\cal
A}^*_{\alpha}(t'\!, {\bf x}) e\!^{- i \epsilon\!_{t\!h}\!(t'\!-\!t)}
F(t'\!-\!t)\hspace{0.1cm} \label{main}
\end{eqnarray}
where ${\cal A}_{\alpha} (t, x) = \sum_i U_{\alpha i } \psi_{i}(t,
x)$ is the probability amplitude corresponding to the flavor
$\alpha$.  The probability in Eq. (\ref{main}) can be explicitly
computed for any type of oscillating particle, provided we specify
its initial state.

To this end, we consider the evolution of an initial wave-packet of
$A$ particles. We assume that the solid angle connecting the
production region and the detection region is very small, so that
only particles with momentum along the axis that connects the two
regions are detected. Let the initial state of the $A$ particles be
a flavor superposition of different Gaussian mass-eigenstates
\begin{eqnarray}
\psi_{i0}( x) =  U^*_{\beta i} \frac{1}{(\pi \sigma^2)^{1/4}}
e^{-\frac{x^2}{2\sigma^2} + i p_{i} x},
\end{eqnarray}
where $\sigma$ is the spread of the initial wave-packet and $\beta$
the initial flavor. Then,
\begin{eqnarray}
{\cal A}_{\alpha}(t, x) = \sum_{i} U^*_{\beta i}U_{\alpha i} (4 \pi
\sigma^2)^{1/4} \nonumber \\
\times \int \frac{dp}{2\pi} e^{- \frac{\sigma^2}{2} (p - p_{i})^2 +i
p x - i E_{i}(p) t - \Gamma_{i}(p)t}, \label{AA}
\end{eqnarray}
where $E_{i}(p) = \sqrt{m_{i}^2 + p^2}$ and $\Gamma_{i}(p)$ are the
decay rates in  the different mass eigenspaces. We employ a commonly
used approximation: we expand $E_{i}(p)$ to first order in $p -
p_{i}$ and $\Gamma_{i}(p)$ to zero-th order in $p- p_{i}$, i.e.,
\begin{eqnarray}
E_{i}(p) = E_{i} + v_{i} (p - p_i); \hspace{1cm} \Gamma_{i} (p) =
\Gamma_{i}, \label{approx}
\end{eqnarray}
where $E_{i} = E_{i}(p_{i})$, $v_{i} = p_{i}/E_{i}$, and $\Gamma_{i}
= \Gamma_{i}(p_{i})$ \footnote{Keeping higher order terms in the
expansion would provide a better approximation that would also
incorporate the effects of wave packet dispersion.}. Then,
\begin{eqnarray}
{\cal A}_\alpha(t, x) = \sum_{i}U^*_{\beta i}U_{\alpha i}\;
\frac{e^{-\frac{(x - v_{i}t)^2}{2 \sigma^2} + i p_{i}x - iE_i t -
\Gamma_{i}t}}{(\pi \sigma^2)^{1/4}}. \label{ampl}
\end{eqnarray}
Substituting (\ref{ampl}) in Eq. (\ref{main})  and letting $T
\rightarrow \infty$, we obtain the probability density for the
detection of flavor $\alpha$ at $x = L$
\begin{eqnarray}
p_{\alpha}(L) = K' \left( \sum_{i} S_{i}  e^{-2
\frac{\Gamma_{i}}{v_i}L} + \sum_{i < j} 2 e^{-
(\frac{\Gamma_{i}}{v_i} +
\frac{\Gamma_j}{v_j})L}\right. \nonumber \\
 \left. \times  Re (T_{ij} \;  e^{
 i k_{ij}
L}) \right), \label{fin}
\end{eqnarray}
where $K' >0$ is a redefined normalization constant, $S_i \sim
|U^*_{\beta i}U_{ \alpha i}|^2$, $T_{ij} \sim U^*_{\beta i}U_{
\alpha i} U_{ \beta j} U^*_{\alpha j}$ (their precise form is not
necessary for the arguments) and
\begin{eqnarray}
 k_{ij} &=& \frac{ E_i
 - \epsilon_{th}}{v_i} - \frac{ E_j
 - \epsilon_{th}}{v_j} - (p_i - p_j). \label{k1}
\end{eqnarray}
If $\epsilon_{th}/E_i << 1$ then  $k_{ij} = m_i^2/p_i - m_j^2/p_j$,
i.e., for $p_i = p + O(m_i)$,  $k_{ij}$ is twice the standard result
$(m_i^2 - m_j^2)/(2p)$.

It is important to emphasize that this result does not depend on any
{\em ad-hoc} assumptions about the initial state. It follows from
Eq. (\ref{prob}), which is valid for a large class of measurements,
and it provides the probabilities for alternatives when the
detection time is not measured. Moreover, Eq. (\ref{k1}) is
insensitive to the approximations we made in this paper. It holds
for
 {\em any} calculation, in which the summation over detection time is performed at the level of
 amplitudes. This is the reason why it
it is discerned even in the simplified plane wave description.

If we do not assume that $\epsilon_{th}/E_i << 1$, Eq. (\ref{k1})
becomes $k_{ij} = (1  - \frac{\epsilon_{th}}{2E})(m_i^2 - m_j^2)/E$,
in the ultra-relativistic limit.  The oscillation wavelength carries
a strong dependence on the threshold energy. This dependence does
not arise {\em at all} in the standard treatment and it is not an
artifact of any approximation. In particular, the presence of an
energy threshold is equivalently described  as a constant potential
$V = - \epsilon_{th}$, in which the particle propagates prior to
detection. The incorporation of this potential to the Hamiltonian
would lead to Eq. (\ref{fin}) even in the simplified plane wave
description.

As a final remark, we note that the kernel $F(t-t')$ in Eq.
(\ref{main}) turns out to cause significant suppression of
interferences in detection time for the amplitude (\ref{ampl}). One
might, however, inquire whether it is possible that in a different
model one might obtain
 $F(t-t') \sim \delta(t-t')$, so that
$p_{\alpha}(L) \sim \int_0^T dt |{\cal A}(t, L)|^2$; the standard
expression for the oscillation wavelength would then follow instead
of (\ref{k1}). However, this would require the presence of an
unusual and  highly efficient decohering mechanism for the detection
time that would be effective in timescales shorter
 than the shortest one  appearing in
(\ref{ampl}), namely, $E_i^{-1}$.  Moreover, such  a  mechanism
would have to be {\em extrinsic} to the physics of particle
oscillations, and it would  have to be postulated {\em ad hoc}.


\begin{thebibliography}{}
\bibitem{osc} M. Gell-Mann and A. Pais, Phys. Rev. 97,  1387 (1955);
B.Pontecorvo, JETP 33, 599 (1957); B. Kayser, Phys. Rev. D24, 110
(1981); C. Giunti, C. W. Kim, J. A. Lee and U. W. Lee,
 Phys. Rev. D48, 4310 (1993). See also the reviews \cite{Zra, Beu03} and
 references therein.

\bibitem{Zra} M. Zralek, Acta Phys. Polon. B29, 3925 (1998).

\bibitem{Beu03} M. Beuthe, Phys. Rept. 375, 105 (2003).

\bibitem{Giu03} C. Giunti, Phys. Scr. 67, 29 (2003).

\bibitem{Lip} H. Lipkin, Phys. Lett. B 642, 366 (2006).

\bibitem{QMT} E. B. Davies, {\em Quantum theory of open systems}
(Academic Press, 1976); P. Busch, P. J. Lahti and P. Mittelstaedt,
{\em The Quantum Theory of Measurement}, (Springer, 1996).


\bibitem{AS} C. Anastopoulos and N. Savvidou,  J. Math. Phys. 47, 122106
(2006);  48, 032106 (2007); 49, 022101 (2008); C. Anastopoulos,  J.
Math. Phys. 49, 022103 (2008).


\bibitem{scg} Of direct relevance is the notion  of "spacetime coarse-graining"; see, J.
B. Hartle, Phys. Rev. D44, 3173 (1991);
 in   Proceedings of the 1992 Les Houches School, Gravitation and
Quantisation, 1993.

\bibitem{SD81} M. D. Srinivas and E. B. Davies, J. Mod. Opt. 28, 981
(1981); 29, 235 (1982).

\bibitem{Sav} For an elaboration of this distinction, see, N.
Savvidou, in {\em Approaches to Quantum Gravity}, ed. D. Oriti
(Cambridge University Press, 2009), and references therein.



\end{thebibliography}
\end{document}